\documentclass[%
 reprint,
 amsmath,amssymb,
 aps,
]{revtex4-1}

\usepackage{graphicx}
\usepackage{dcolumn}
\usepackage{bm}
\usepackage[colorlinks=true]{hyperref}
\usepackage{siunitx}
\usepackage{subfigure}

\usepackage{xcolor}
\usepackage{soul}
\usepackage{systeme}

\newcommand{\appendixsectionname}{\appendixautorefname}
\newcommand{\amended}[1]{\color{black}#1\color{black}}
\begin{document}


\title{Relation between charging times and storage properties of nanoporous supercapacitors}
\author{
Timur Aslyamov
\\
\texttt{t.aslyamov@skoltech.ru}
\\
Center for Design, Manufacturing and Materials, Skolkovo Institute of Science and Technology, Bolshoy Boulevard 30, bld. 1, Moscow, Russia 121205
\\
Konstantin Sinkov\\
\texttt{sinkovk@gmail.com}\\
Schlumberger Moscow Research, Leningradskoe shosse 16A/3, Moscow, Russia 125171\\
Iskander Akhatov\\
Center for Design, Manufacturing and Materials,
Skolkovo Institute of Science and Technology,
Bolshoy Boulevard 30, bld. 1, Moscow, Russia 121205
}
\date{\today}

\begin{abstract}

Investigating the correlations between dynamic and static storage properties of nanoporous electrodes is beneficial for further progress of supercapacitors-based technologies. While the dependence of the capacitance on the pores' sizes is well described by classical Density Functional Theory (c-DFT), the lack of dynamic c-DFT extension capable for correct estimation of the charging time has been noted in the literature. Here, we develop a dynamic model of the electrolyte inside nanopores based on c-DFT and realistically describing both the time-dependent charging process and maximum static capacitance. Our calculations show that the charging starts with a square-root dependency of the total charge on time and then follows two subsequent exponential trends with significantly different time scales that agree with published simulations. We demonstrate that the full charging time corresponds to the timescale of either the first or the second exponential trend depending on the pores' size. Also, we find analytical expressions to fit the timescales for a wide range of parameters. Derived correlations provide the relation of charging time to pores' size, applied voltage, and final ions' densities inside the pore, making these expressions useful to design supercapacitors with an optimal combination of power and energy characteristics.   
\end{abstract}

\maketitle
Among all modern energy sources, the supercapacitors demonstrate an extraordinary power density and an extremely long cycling life \cite{simon2010materials}. Such rapid charging and discharging performance results from the fast adsorption-electrostatic processes, making the supercapacitor technology ecology friendly. These advantages open a wide range of the possible applications from small devices \cite{simon2020perspectives} to electrocars \cite{horn2019supercapacitors}. The wide distribution of the supercapacitors technology is limited by the relatively low energy density \cite{shao2020nanoporous}. The nanoporous electrodes' implementation has led to the serious enhancement of the energy density due to the significant capacitance increase first experimentally observed in subnanoporous carbon materials \cite{chmiola2006anomalous}. Moreover, the later experiments have shown optimal (yielding the highest capacitance) pore size approximately corresponding to electrolyte's molecular diameter \cite{largeot2008relation}. 
The capacitance's oscillatory behavior as a function of pore size has been successfully described in terms of the classical Density Functional Theory (c-DFT) \cite{jiang2011oscillation}, accounting for confined properties of the charged hard spheres at applied electrostatic potential. Modern state of c-DFT approach \cite{hartel2017structure} allows to investigate how the supercapacitors' parameters, namely electrodes' pore sizes \cite{pizio2012electric,jiang2013microscopic} and electrolyte composition \cite{lian2016enhancing, neal2017ion, osti2018mixed}, affect energy storage performance.

Despite the long history study \cite{delevie1963porous, delevie1964porous, posey1966theory} of the porous electrodes' charging, the overwhelming part of the existing dynamic models correspond to meso- and macro-pores (pore size $H\geq\SI{2}{nm}$), 
where the properties of the confined dilute electrolytes are similar to the bulk ones. 
In this case, the charging time is often estimated by the linearization of Poisson-Nernst-Planck (PNP) equations \cite{bazant2004diffuse} referred to as {\it RC} Transmission Line Model (TLM), first proposed and thoroughly studied in \cite{delevie1963porous, delevie1964porous}. This approach provides an equivalent circuit of the linear resistors and capacitors to account for bulk and Electric Double Layer (EDL) states of the electrolyte, respectively. 
Besides the assumption that the pores size is much larger than Debye length $H \gg \lambda_\text{D}\sim\SI{1}{nm}$, the linear TLM is derived for the extremely low applied potential $U$ such that $U \ll k_B T/e$, where $k_B$ is Boltzmann constant, $T$ is the temperature and $e$ is the electron charge. 
Because of the latter assumption, TLM misses certain critical phenomena \cite{biesheuvel2010nonlinear, mirzadeh2014enhanced} at the conditions corresponding to the majority of applications in technology ($U\sim 1-100 k_B T/e$). 
Very recently, TLM approach has been applied to nanoporous electrodes ($H\geq 2\lambda_D$), and realistic scaling of the charging time has been obtained in terms of physically determined parameters \cite{lian2020blessing, janssen2021transmission}. Authors of \cite{lian2016enhancing} also have shown that two comparable timescales exist at high applied potentials: the first one corresponds to the equivalent circuit model, and the second timescale is related to the adsorption process. The detailed molecular dynamics (MD) simulations \cite{kondrat2010superionic,breitsprecher2018charge,breitsprecher2020speed} demonstrate that the second timescale depends on the confined fluid properties, which can be reproduced correctly neither by linearized TLM nor by a more general PNP approach. 

In this work, we describe charging dynamics in the nanopores of width until the molecular size. Such sub-nanoporous electrodes are extremely important for the further progress of supercapacitors-based technologies. However, sub-nanoscale confinement induces the significant slow down of the ions' movement and, as a result, the charging. Our theory describes three consequent charging regimes: the initial root-square process and two exponential regimes with notably different time scales. The last exponential regime plays a crucial role in the ultra-narrow pores and results in the charging slow down. Before the current work, the details of the charging inside ultra-narrow pores has been investigated using MD simulations only \cite{kondrat2014accelerating, breitsprecher2018charge, breitsprecher2020speed}. Our theory accounts for the realistic thermodynamic and packing properties of the finite-size ions inside nanopores that is the advantage over recent published dynamic models \cite{lian2020blessing, gupta2020charging} and allows us to relate the charging times with the predetermined nanoporous supercapacitors characteristics.

\begin{figure}
    \centering
    \includegraphics[width=0.45\textwidth]{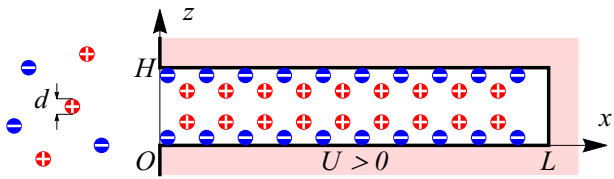}
    \caption{Slit pore connected to bulk volume of symmetric electrolyte, constant electric potential is applied to the pore walls.
    }
    \label{fig:sketch}
\end{figure}

\amended{
We use the slit pores geometry shown in Fig.~\ref{fig:sketch}, which describes the spatial structure of the modern porous materials such as graphene based elecrtodes \cite{yoo2011ultrathin, yang2013liquid} and graphene oxide (GO)/MXene fibers \cite{he2020effects, li2021assembly} exhibiting aligned slit-nanopores of width $H$ comparable to electrolyte diameter $d$ and length $L \gg H$. 
These materials show promising applications for wearable electronics and smart textiles devices due to outstanding energy density and high flexibility \cite{yang2017mxene}. 
Also, a model of individual slit nanopore considered in this work can be used as a building block of pores-network in the case of nontrivial pores size distribution \cite{vasilyev2019connections, lian2019non}.
Moreover, our geometrical representation of the pore closely resembles recent molecular dynamic simulations of the supercapacitors \cite{breitsprecher2018charge, breitsprecher2020speed}, which allows us to make a comparison of our predictions with the MD-simulated electrolyte's behaviour.
}
Slit pore geometrical constraints induce the structured ions packing that corresponds to the density distribution functions instead of bulk homogeneous density used in macro-scale models. In the case of slit pore geometry, the dynamic component density $\rho_i(t, x, z)$ depends on the coordinate $x$ in lateral direction along the pores surface and the normal distance to the surface $z$. To describe dynamics of the charging process, it is possible to extend c-DFT approach to time-dependent version \cite{jiang2014time}, defined in the general form as 
\begin{equation}
    \label{eq:transport}
    \partial_t \rho_i - \beta D_i \nabla \left(\rho_i \nabla \varphi_i\right) = 0,
\end{equation}
where $D_i$ is the diffusion coefficient of $i$-th component, $\beta = 1 / k_B T$, $\nabla=\{\partial_x, \partial_z\}$ is 2D gradient vector, $\varphi_i(t,x,z)$ is the local electrochemical potential of the $i$-th component. 

We use one of the most popular c-DFT approaches based on confined hard sphere model \cite{roth2010fundamental} and electrostatic extension \cite{wang2011weighted} accounting for the contributions from Coulomb interaction and additional finite size residual correlations. The electrode's pores are considered as the open system connected with the bulk electrolyte. The confined density distributions are described in terms of c-DFT from the grand potential $\Omega$ minimization $\delta\Omega/\delta\rho_i=0$. In accordance with \cite{jiang2014time} the potentials $\varphi_i(t,x,z)$ have the following form
\begin{equation}
\label{eq:potential_definition}
\varphi_i = k_\text{B}T\log\left(\rho_i\Lambda^3\right)+e Z_i \psi +\frac{\delta f_\text{exc}}{\delta \rho_i}.
\end{equation}
The electric potential $\psi(x,z,t)$ satisfies the Poisson equation 
\begin{equation}
\label{eq:Poisson_equation}
\beta e \Delta\psi=-4 \pi l_\text{B} \sum_{i=1}^n Z_i\rho_i,
\end{equation}
where $\Delta = \partial_{xx} + \partial_{zz}$ is the 2D Laplace operator and $l_\text{B}=\beta e^2/4\pi \epsilon \epsilon_0$ is the Bjerrum length. The detailed description of used c-DFT approach including particular form of the excess term $f_\text{exc}$ can be found in \cite{supplementary}. 
As one can see from \cite{pizio2012electric} implemented static DFT model \cite{supplementary} describes the realistic behaviour of the capacitance's properties.

Initially, no electrostatic potentials are applied that corresponds to the zero total charge of the symmetric electrolyte. We consider the step-like charging, when the potential turns on abruptly from zero to some positive value $U>0$. 
\amended{
The electrode's charge is controlled by the distribution of the confined electrolyte in the electrostatic field. 
More precisely, the applied electrostatic potential 
induces the influx of the oppositely charged ions (counter-ions, negative in the case of $U > 0$) to the pores. In contrast, the ions with the charge of the same sign as applied potential (co-ions, positive for $U > 0$) are pushed out from the porous volume. Inside sub-nanopores, local increase of the counter-ions density can strongly influence the co-ions desorption, hindering their release due to counter-ions clogging and thus leading to the notable slowdown of the charging process \cite{breitsprecher2018charge}.
}
The main variables of interest are the charges associated with individual electrolyte components ${Q_i (t) = -e Z_i \int^L_0 \int^H_0 dx \: dz \: \rho_i (t, x, z)}$ and the total charge density $Q(t) = \sum^n_{i=1} Q_i(t)$. The external potential $U$ induces the charging to the final charge $Q_\infty$ during a time $\tau$ depending on the external parameters and inner structure of the electrolyte. The final state corresponds to the steady distribution $\rho^\infty_i(t,x,z)=\rho^\infty_i(z)$, which is in the equilibrium with the bulk electrolyte. 

Due to slit pore geometry, the timescale of densities relaxation to the equilibrium distribution in the transverse direction $z$ is much less than characteristic timescale of transport along the lateral coordinate $x$. Exploiting this fact and performing scaling analysis for the equations \eqref{eq:transport} and \eqref{eq:Poisson_equation}, we derive asymptotic 1D model describing charging dynamics in terms of the pore cross-section averaged quantities \cite{supplementary}. The asymptotic model approximates the original 3D model in the leading order of the $H^2 / L^2 \ll 1$ parameter. The resulting transport equations has the following form:
\begin{equation}
    \label{eq:transport_average}
    \partial_t \overline{\rho}_i - \beta D_i \partial_x \left(\overline{\rho}_i \partial_x \overline{\varphi}_i\right) = 0.
\end{equation}
Here, $\overline{\rho}_i (t,x) = \int^H_0 dz \: \rho_i(t,x,z) / H$ are pore cross-section-averaged densities and the averaged potentials are functions of averaged densities only $\overline{\varphi}_i = \overline{\varphi}_i(\overline{\rho}_1, ..., \overline{\rho}_n)$. The $\overline{\varphi}_i(\overline{\rho}_1, ..., \overline{\rho}_n)$ dependencies are defined by the cross-section-wise solution of c-DFT model for confined density distributions \cite{supplementary}. The charges of electrolyte components are conveniently defined in terms of average densities as $Q_i(t) = -e Z_i H \int^L_0 dx \: \overline{\rho}_i (t, x)$ and the final steady state of full charging corresponds to $\overline{\rho}_i (t, x) = \text{const}$.

\begin{figure*}[t!]
    \includegraphics[width=\textwidth]{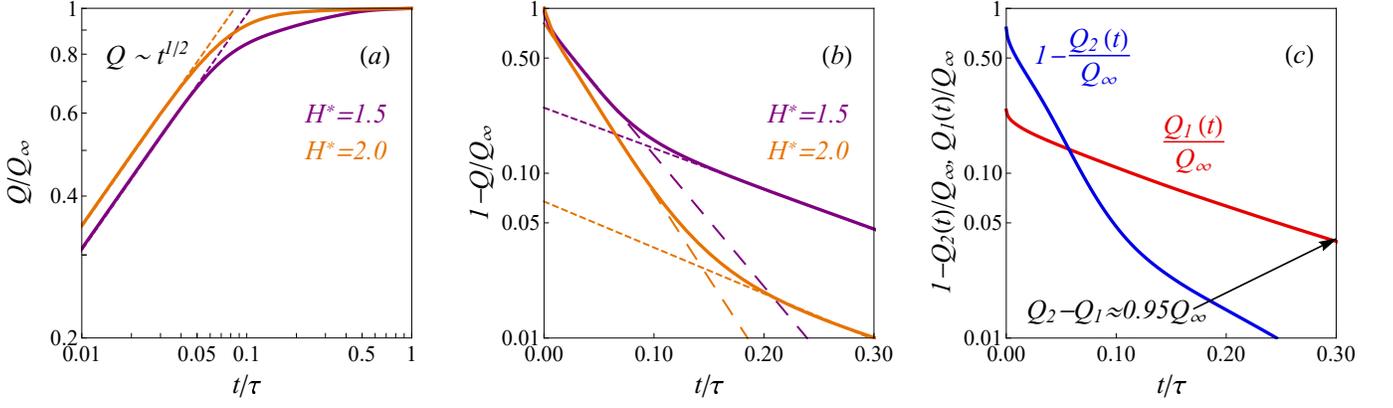}
    \caption{
    (a) The total charge (solid lines) follows square-root trend (dashed lines) at early times. (b) The total charge (solid lines) follows two exponential trends (dashed and dotted lines) at middle and late times. (c) The charge contribution of co-ions (red) is a notably slower function of time than the contribution of counter-ions (blue).}
    \label{fig:charge_profile}
\end{figure*}

Similarly with the realistic simulations \cite{breitsprecher2018charge, kondrat2014accelerating}, we assume that electrolyte is a symmetric two component mixture  consisting of the molecules with the diameters $d_1=d_2=d$ and the charge valences $Z_1=-Z_2=1$. For the sake of simplicity, we set the diffusion coefficients of the components to be equal $D_1=D_2=D$. In the case of the symmetric electrolyte, it is also useful to introduce, following \cite{pizio2012electric}, the dimensionless variables $H^*=H/d$, $U^*=e U/k_B T$, $Q^*=Q d^2/e$, $\rho^* = \rho d^3$. 
It can be shown from dimensional arguments that the characteristic time for the considered problem is $L^2 / D$ and one can also introduce the scaled time $t^* = t D / L^2=t/\tau$. 

An example of the calculated time-dependent charges $Q(t)$ for subnanopore electrodes (${H^*=1.5}$ and ${H^*=2}$) at sufficiently high potential ${U^*=10}$ is shown in Fig.~\ref{fig:charge_profile}(a-b). As one can see from Fig.~\ref{fig:charge_profile}(a), the root-square law $Q\sim\sqrt{t}$ describes the notable part of the charging process at early times well. However, when the charge $Q$ approaches saturation $Q_{\infty}$, the trend changes to the exponential one. Fig.~\ref{fig:charge_profile}(b) shows that in the case of larger pores with $H^*=2$, the charging up to almost $95\%$ is described by the following equation: 
\begin{equation}
\label{eq:charge_late_time}
\frac{Q}{Q_\infty}=1-\frac{8}{\pi^2} e^{-t/\tau_1}.
\end{equation}
Expression \eqref{eq:charge_late_time} is the leading term of the analytical solution of the TLM \cite{mirzadeh2014enhanced}. Despite that the pores sizes $H \sim d$ and potentials $U> k_\text{B}T/e$ are significantly beyond the ranges of TLM applicability, the published computer simulations \cite{kondrat2014accelerating, bi2020molecular} show the adequacy of exponential trend \eqref{eq:charge_late_time} for fitting of the charge dependency on time. 
We observed that at late times the calculated profiles $Q(t)$ in Fig.~\ref{fig:charge_profile}(b) follow another exponential trend succeeding \eqref{eq:charge_late_time}, which is notably slower and describes the charging until almost full saturation. 

\amended{
The predicted in our work three consequent dynamic regimes describe the results of the molecular dynamics simulations published in  \cite{kondrat2014accelerating, breitsprecher2018charge, breitsprecher2020speed}. 
Similarly with \cite{kondrat2014accelerating, breitsprecher2018charge} we observed a stark contrast between the charging dynamics inside ultra-narrow  $H^*=1.5$ and wider $H^*=2$ pores shown in Fig.~\ref{fig:charge_profile}. 
In Fig.~\ref{fig:charge_profile}(b) the first exponential regime \eqref{eq:charge_late_time} in pores $H^*=1.5$ covers the charging up to only $75\%$. Therefore, the significant part of the full charging inside ultranarrow pores ($H^*=1.5$) is defined by the slow-downed second exponential trend: the charge to $95\%$ is $\sim 2$ times slower than in the case of wider pores $H^*=2$. 
We considered the dynamics of the co-ions/counter-ions separately to explain this phenomenon arisen inside ultra-narrow pores. %
Our calculations shown in Fig.~\ref{fig:charge_profile}(c) demonstrates that the contribution to the total charge from the co-ions $Q_1(t)$ is a significantly slower function of time than the counter-ions contribution $Q_2(t)$. Therefore, the total charge to the final value $Q=Q_2-Q_1=0.95Q_\infty$ demands such a long time due to the co-ions' slow-release (desorption). This phenomenon is confirmed by MD simulations of the symmetric electrolyte \cite{breitsprecher2018charge}, which revealed that co-ions are trapped in the crowded contour-ion phase.
}

\amended{
The experimentally observed charging time for the supercapacitors is around $10^3~\SI{}{s}$, while the published dynamic DFT models result in the enormously underestimated values $\sim 10^{-9}~\SI{}{s}$  \cite{lian2020blessing}. The proposed model provides both the realistic behaviour of the capacitance’s properties (\cite{supplementary}, section IV) and correct charging-time scales. To provide numerical estimations we have used experimental parameters from works \cite{janssen2017coulometry} and \cite{lian2020blessing}: the pore length is $L=\SI{0.5}{mm}$ and confined diffusion coefficient $D=2\times10^{-10}~\SI{}{m^2s^{-1}}$. For example, our calculations for the pores $H^*=1.5$ and $H^*=2$ shown in Fig.~2(b) correspond to the full charging time around \SI{700}{s} and \SI{400}{s}, respectively, that agrees with the experimental characteristics. 
}

\begin{figure}
    \centering
    \includegraphics[width=0.45\textwidth]{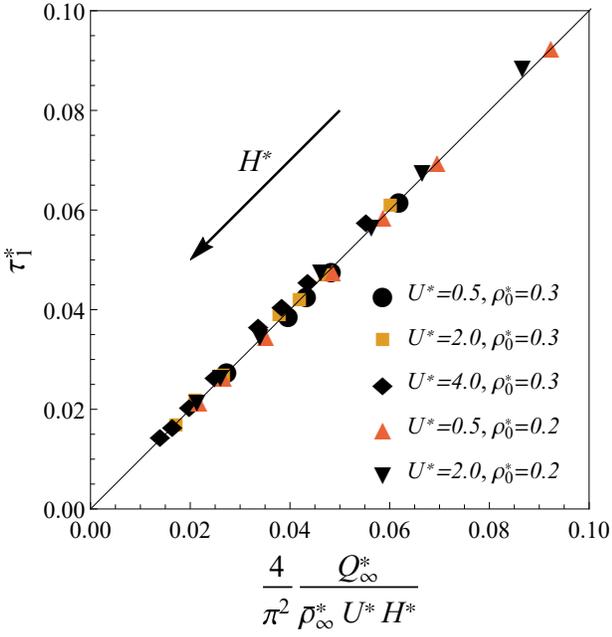}
    \caption{The calculated dimensionless charging time $\tau_1^*$ for the electrodes with $H^*>2$ at the external potentials $U^*<4$ and the bulk electrolyte densities $\rho^*_0$ versus the scaling law \eqref{eq:scaled_time}. The arrow shows direction of the pores size increase.}
    \label{fig:scaled_time_charging_1}
\end{figure}

The calculations for the wide range of the parameters show that the full charging (say 95\% for the sake of concreteness) in pores with $H^*\geq 2$ is described by the first exponential trend \eqref{eq:charge_late_time}. For this reason, we use expression \eqref{eq:charge_late_time} to fit the charging time $\tau^*_1$ from the calculated profiles $Q(t)$. Fig.~\ref{fig:scaled_time_charging_1} demonstrates that the charging time for the electrodes of width $2\leq H^* \leq 4$ and various electrolyte bulk densities $\rho_0$ and applied potentials $U^*\leq 4$ can be explicitly expressed in terms of macroscopic parameters as
\begin{equation}
\label{eq:scaled_time}
\tau_1^* =\frac{4}{\pi^2}\frac{Q^*_{\infty}}{\overline{\rho}^*_{\infty}U^*H^*}.
\end{equation}
Here, $\overline{\rho}^*_\infty = \overline{\rho}^*_{1,\infty} + \overline{\rho}^*_{2,\infty}$ is the total final density. The potentials considered here  (up to $\SI{0.1}{V}$ in dimensional terms) correspond to lower range of the values used in the modern experiments \cite{evlashin2020role, janssen2017coulometry, prehal2018salt}, but are significantly beyond the formal applicability range of the TLM \cite{mirzadeh2014enhanced}. The coefficient $4/\pi^2$ is obtained from the analysis of expression \eqref{eq:scaled_time} in the limit of large pores ($H^* \gg 1$) at extremely low potentials ($U^* \ll 1$), where the charging can be described by TLM \cite{mirzadeh2014enhanced}. In this limit, the confined fluid density tends to the bulk value $\overline{\rho}\to\rho_0^*$ and the charge can be calculated from the linearization of the Gouy-Chapman theory $Q\simeq e \rho_0 \lambda_D U^*$. Substituting these approxamiations in expression \eqref{eq:scaled_time}, we obtain  $\tau^*_\text{TLM}=4\lambda_\text{D} / (\pi^2 H)$, which is the relaxation time of the leading term in TLM analytical solution \cite{mirzadeh2014enhanced}.  
\amended{
Also,  we observed an interesting connection between expression \eqref{eq:scaled_time} and the very recent result for the charging time-scale published in \cite{bi2020molecular}. As one can see from section V of \cite{supplementary} the application of adopted TLM expression from \cite{bi2020molecular} to slit pores gives the following expression for charging time $\tau_\text{aTLM}=4 Q_\infty L^2/\pi^2\sigma H U$. Here $\sigma$ is the ionic conductivity inside the pore affected by the confined properties of the electrolyte. We express $\sigma$ in terms of the electrical mobility of ions $m = e D / k_B T$ and ions density $\overline{\rho}_\infty$ as $\sigma = e \overline{\rho}_\infty m$, which leads to the equality of $\tau_\text{aTLM}$ from \cite{bi2020molecular} and $\tau^*_1$ given by \eqref{eq:scaled_time}.
}
Thus, our result \eqref{eq:scaled_time} generalizes the TLM predictions for the case of the charging nanoporous supercapacitors at higher potentials. 

\begin{figure}[t!]
    \centering
    \includegraphics[width=0.45\textwidth]{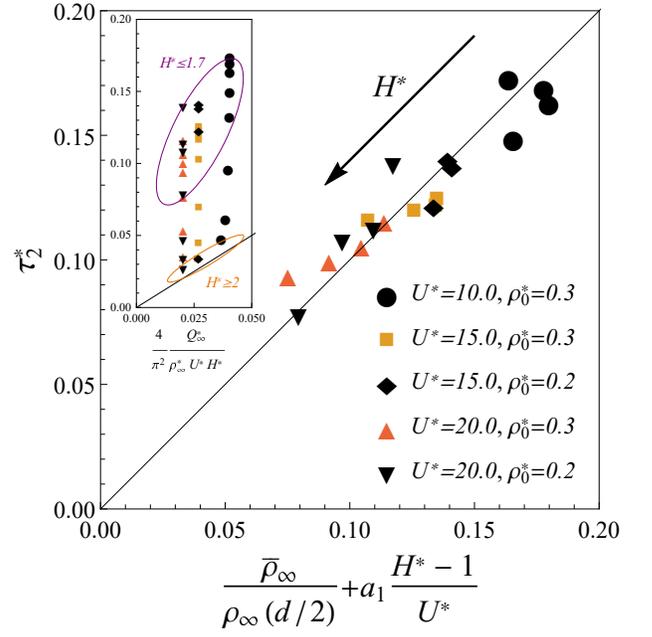}
    \caption{
    The dimensionless charging timescale $\tau^*_2$ near full charging for the electrodes with $1.3 \leq H^*\leq 1.7$ at the high external potentials $10 \leq U^* \leq 20$ versus the scaling law \eqref{eq:charging_time_2}. The inset shows the same timescale versus the scaling law \eqref{eq:scaled_time}; the pores with $H^* \leq 1.7$ and $H^* \geq 2.0$ are marked with purple and orange ellipses.
    }
    \label{fig:scaled_time_charging_2}
\end{figure}

As discussed above, in the more narrow pores ($H^*<2$) the second exponential regime of charging can influence the full charging time. To estimate this effect numerically, we described the charge profiles near the saturation $Q(t)\sim 0.95Q_\infty$ in terms of another exponential trend $Q / Q_\infty=1-A_2 e^{-t/\tau_2}$, where $\tau_2\geq\tau_1$ and $A_2$ are the fitting parameters. The results for the corresponding relaxation time $\tau_2^*$ are shown in inset of Fig.~\ref{fig:scaled_time_charging_2}. 
As one can see from this inset, expression \eqref{eq:scaled_time} fits the charging time in larger pores ($H^*\geq 2$). But we observed that in the case of the ultranarrow pores $H^*\leq 1.7$ the charging slow down becomes notable.
We observed that the charging slow down starts as the pores width becomes less than two molecular diameters ($H^*< 2$). 
In such confinement, the electrolyte behaviour near the wall crucially influence on the inner structure. Accordingly, it is reasonable to correlate the charging inside narrow pores not only with the average inner density $\overline{\rho}$ but the wall-contact density as well.
As one can see from Fig.~\ref{fig:scaled_time_charging_2}, the charging time in ultranarrow pores ($H^*\leq 1.7$) at high potential can be fitted to the following expression
\begin{equation}
\label{eq:charging_time_2}
\tau^*_2=\frac{\overline{\rho}_\infty}{\rho_{\infty}(d/2)}+a_1\frac{H^*-1}{U^*}
\end{equation}
where $\rho_{\infty}(d/2)=\rho_{\infty,1}(d/2)+\rho_{\infty,2}(d/2)$ is the wall-contact density at the final state of complete charging and ${a_1\simeq 0.6}$ is the fitting parameter. The inverse dependence of charging time \eqref{eq:charging_time_2} on the potential $U$ contributes as the pore's width increases. This behaviour is consistent with the high potential limit ($H^*\overline{\rho}^*_\infty\simeq Q^*_\infty$) of expression \eqref{eq:scaled_time}, which fits well the data around $H^*=2$. 
In the ultra-narrow pores ($H^*\to 1$), where the final co-ions density is much less than counter-ions density $\overline{\rho}_{\infty,1} \ll \overline{\rho}_{\infty,2}$, expression \eqref{eq:charging_time_2} can be written as $\tau^*_2=\overline{\rho}_{\infty,2} / \rho_{\infty,2}(d/2)$. Then slow desorption of co-ions corresponds to the diffusion process with the effective diffusion coefficient $D_1=\overline{\rho}_{\infty,2}D/\rho_{\infty,2}(d/2)$, which is defined by the density and structure of the counter-ion component distribution. This result confirms the idea that the abrupt counter-ions adsorption induces slow diffusion of the co-ions.

In conclusion, we developed the model of the charging dynamics in nanopores accounting for the confined properties of the electrolyte. 
The predictions of our theory agree with published simulations and experimental data.  
Another striking result is two analytical expressions for the time scales of the exponential regimes, which depend on the predetermined supercapacitors parameters (size's characteristics, applied potential, and electrolytes density distribution at saturated state calculated from c-DFT or MD). We identified threshold pores width, below which the second exponential regime defines the full charging time. In this case, the charging inside ultra-narrow pores is notably slower and depends on contact wall electrolytes density. These numerical estimations are crucial to avoid the power density decrease using the ultra-narrow pores. 
Thus, the developed model will help investigate the relations between the supercapacitors' storage and dynamics properties.

\begin{acknowledgments}
T.A. acknowledges the financial support from the Russian Science Foundation (project number: 20-72-00183). K.S. is grateful to Schlumberger management for the permission to publish this work. The authors are grateful to Mikhail Stukan for useful comments. 
T.A. and K.S. contributed equally to this work. 
\end{acknowledgments}
\bibliography{supercapacitors_DDFT}
\newpage
\onecolumngrid
\appendix
\section{Derivation of pore cross-section averaged equations}
Here we describe derivation of pore cross-section averaged equations \eqref{eq:transport_average} from the original three dimensional system \eqref{eq:transport}, \eqref{eq:potential_definition}, \eqref{eq:Poisson_equation}. We consider transport along the slit pore aligned with $x$-axis and transverse to $z$-axis. Aspect ratio of the pore is assumed to be large, and accordingly, we adopt classical thin-film~/~lubrication approximation scaling. The pore width $H$, length $L$, and applied voltage $U$ are used as scales for $z$ coordinate, $x$ coordinate and potentials $\varphi_i$, respectively. Additionally, for the sake of concreteness we scale $D_i$ for different components with the value of the first component diffusion coefficient $D$ at bulk conditions. Timescale of the problem is defined by $L^2 / D \beta U$ and characteristic density is $\beta e U / l_B H^2$. 

After scaling, the dimensionless transport and Poisson equations are
\begin{equation}
    \label{eq:dimensionless_transport}
    \delta \partial_t \rho_i - \delta \partial_x \left(K_i \partial_x \varphi_i\right) - \partial_z \left(K_i \partial_z \varphi_i\right) = 0,
\end{equation}
\begin{equation}
    \label{eq:dimensionless_Poisson}
    \delta \partial_{xx} \psi + \partial_{zz} \psi - \sum^n_{i=1} Z_i \rho_i = 0.
\end{equation}
Here, $\delta = H^2 / L^2 \ll 1$ and additional definition $K_i~=~D_i \rho_i$ is introduced. 

The system is supplied with boundary conditions enforcing zero fluxes
\begin{equation}
    \label{eq:dimensionless_bc_transport}
    -K_i \partial_z \varphi_i = 0, \quad z = 0, 1.
\end{equation}
and the value of electrostatic potential
\begin{equation}
    \label{eq:dimensionless_bc_Poisson}
    \psi = 1, \quad z = 0, 1.
\end{equation}
at the channel walls.

We seek formal asymptotic expansion of densities $\rho_i$ and potential $\psi$ in power series of $\delta \to 0$ 
\begin{eqnarray*}
    \rho_i &=& \rho^0_i + \delta \rho^1_i + \dots, \\
    \psi &=& \psi^0 + \delta \psi^1 + \dots.
\end{eqnarray*}

Substituting the latter expansion to \eqref{eq:dimensionless_transport}, \eqref{eq:dimensionless_Poisson}, \eqref{eq:dimensionless_bc_transport}, \eqref{eq:dimensionless_bc_Poisson} and collecting terms of the same order, one can obtain

$O\left(1\right)$ {\it problem}:
\begin{equation}
    \label{eq:dimensionless_transport_0}
    - \partial_z \left(K^0_i\partial_z \varphi^0_i\right) = 0,
\end{equation}
\begin{equation}
    \label{eq:dimensionless_Poisson_0}
    \partial_{zz} \psi^0 - \sum^n_{i=1} Z_i \rho^0_i = 0,
\end{equation}
\begin{equation}
    \label{eq:dimensionless_bc_transport_0}
    -K^0_i \partial_z \varphi^0_i = 0, \quad z = 0, 1.
\end{equation}
\begin{equation}
    \label{eq:dimensionless_bc_Poisson_0}
    \psi^0 = 1, \quad z = 0, 1.
\end{equation}
Here, $K^0_i = D_i \rho^0_i$, $\varphi^0_i = \varphi_i \left[\rho^0_1, ..., \rho^0_n,  \psi^0\right]$. Solution of the $O \left(1\right)$ problem will be discussed below. Here we only note that using \eqref{eq:dimensionless_transport_0} and the boundary condition \eqref{eq:dimensionless_bc_transport_0}, one can also promptly get 
\begin{equation}
    \label{eq:dimensionless_constant_potential}
    \partial_z \varphi^0_i = 0.
\end{equation}

Details of $O(1)$ problem are further considered in \appendixsectionname~\ref{sec:o1_problem}.

$O\left(\delta\right)$ {\it problem}:
\begin{equation}
    \label{eq:dimensionless_transport_1}
    \partial_t \rho^0_i - 
    \partial_x \left(K^0_i \partial_z \varphi^0_i\right) - 
    \partial_z \left(K^0_i \partial_z \varphi^1_i + K^1_i \partial_z \varphi^0_i\right) = 0,
\end{equation}
\begin{equation}
    \label{eq:dimensionless_Poisson_1}
    \partial_{xx} \psi^0 + \partial_{zz} \psi^1 - \sum^n_{i=1} Z_i \rho^1_i = 0,
\end{equation}
\begin{equation}
    \label{eq:dimensionless_bc_transport_1}
    -K^0_i \partial_z \varphi^1_i - K^1_i \partial_z \varphi^0_i = 0, \quad z = 0, 1.
\end{equation}
\begin{equation}
    \label{eq:dimensionless_bc_Poisson_1}
    \psi^1 = 0, \quad z = 0, 1.
\end{equation}
Here $K^1_i~=~D_i \rho^1_i$ and
\begin{equation*}
    \varphi^1_i~=~\sum^n_{j=1} \frac{\delta \varphi_i}{\delta {\rho_j}} \left[\rho^0_1, ..., \rho^0_n,  \psi^0\right] \rho^1_j + \frac{\delta \varphi_i}{\delta \psi} \left[\rho^0_1, ..., \rho^0_n, \psi^0\right] \psi^1.
\end{equation*}

Solution of the $O\left(\delta\right)$ problem is beyond the scope of the study. We limit ourselves to consideration of dynamical problem in the leading order of approximation and use $O\left(\delta\right)$ problem for rigorous derivation of the averaged equations only. 

Integrating \eqref{eq:dimensionless_transport_1} over the pore width and using boundary conditions \eqref{eq:dimensionless_bc_transport_1} and corollary \eqref{eq:dimensionless_constant_potential} one can get
\begin{equation}
    \label{eq:dimensionless_transport_average}
    \partial_t \overline{\rho}^0_i - \partial_x \left(\overline{K}^0_i \partial_x \overline{\varphi}^0_i\right) = 0.
\end{equation}
Here, $\overline{f} = \int^1_0 dz f$. The condition \eqref{eq:dimensionless_constant_potential} is used here while integrating the second term of \eqref{eq:dimensionless_transport_1} by parts and to replace $\varphi^0_i$ by $\overline{\varphi}^0_i$. The equation \eqref{eq:transport_average} is essentially \eqref{eq:dimensionless_transport_average} written in dimensional terms after dropping superscripts.
\section{{\it \textbf{O}}(1) problem}
\label{sec:o1_problem}
Casting the equations \eqref{eq:dimensionless_Poisson_0}, \eqref{eq:dimensionless_bc_Poisson_0} and \eqref{eq:dimensionless_constant_potential} back to dimensional variables we get
\begin{equation}
\label{eq:Poisson_0}
\beta e \partial_{zz}\psi^0=-4 \pi l_\text{B} \sum_{i=1}^n Z_i\rho^0_i,
\end{equation}
\begin{equation}
\label{eq:bc_Poisson_0}
\psi^0 = U, \quad z = 0, H,
\end{equation}
\begin{equation}
\label{eq:constant_potential}
\partial_z \varphi^0_i = 0.
\end{equation}

The solution of \eqref{eq:Poisson_0} with boundary conditions \eqref{eq:bc_Poisson_0} can be written in the integral form
\begin{equation}
\label{eq:Poisson_solution}
\beta e \psi^0 = \beta e U+\frac{4\pi l_\text{B} z}{H}\int_0^H dz' (H-z') \sum_{i=1}^n Z_i \rho^0_i
-4\pi l_\text{B}\int_0^z dz' (z-z') \sum_{i=1}^n Z_i \rho^0_i.
\end{equation}

It follows from \eqref{eq:constant_potential} that $\varphi^0_i$ doesn't depend on $z$ coordinate $\varphi^0_i = \overline{\varphi}^0_i (t, x)$. Substituting the latter to \eqref{eq:potential_definition} and rearranging terms, one can get the following equation for density distribution across the pore
\begin{equation}
\label{eq:density_distribution_0}
\rho^0_i = \frac{1}{\Lambda^3} \exp\left(\beta \overline{\varphi}^0_i\right) E^0_i.
\end{equation}
Here, 
\begin{equation*}
E^0_i = \exp\left(-\beta e Z_i \psi^0 - \beta \frac{\delta f_\text{exc}}{\delta \rho_i}[\rho^0_1, ..., \rho^0_n]\right)
\end{equation*}
and further details on $f_\text{exc}$ can be found in \appendixsectionname~\ref{sec:DFT}.

Formal integration of \eqref{eq:density_distribution_0} over $z \in [0, H]$ and the fact that $\overline{\varphi}^0_i = \overline{\varphi}^0_i (t, x)$ allow to eliminate potential $\overline{\varphi}^0_i$ and rewrite the equation in terms of average density $\overline{\rho}^0_i$ as
\begin{equation}
\label{eq:density_distribution_1}
\rho^0_i = \overline{\rho}^0_i \frac{E^0_i}{\overline{E}^0_i}.
\end{equation}

We look for effectively 1D equilibrium densities' distributions depending on $t$ and $x$ only parametrically via the average densities ${\rho^0_i = \rho^0_i (\overline{\rho}^0_i (t, x), z)}$. Once the distributions are found from \eqref{eq:density_distribution_1} and \eqref{eq:Poisson_solution}, one can evaluate \eqref{eq:potential_definition} at any coordinate $z$, and thus, get the potentials as functions of average densities ${\overline{\varphi}^0_i = \overline{\varphi}^0_i (\overline{\rho}^0_i, ..., \overline{\rho}^0_n)}$. 

\section{Notes on numerical solution}
Numerical solution of the transport equations \eqref{eq:transport_average} involves two different tasks: solution of the dynamic equations given the potentials as function of densities and evaluation of potentials itself. 

For given potentials, spatial discretization of the system \eqref{eq:transport_average} is performed on the uniform staggered grid using finite volume method. The resulting system of nonlinear ODEs is solved by the built-in method of Wolfram Mathematica \cite{mathematica2020}.

The potentials as functions of densities are defined by the solution of $O\left(1\right)$ problem described in \appendixsectionname~\ref{sec:o1_problem}. The system of equations \eqref{eq:density_distribution_1} for $i = \overline{1, n}$ comprises the fixed-point problem with the right-hand side defined by the dependency of excess energy variation on densities described in \appendixsectionname~\ref{sec:DFT} and the solution \eqref{eq:Poisson_solution} of the Poisson equation \eqref{eq:Poisson_0}, \eqref{eq:bc_Poisson_0}. Given $\overline{\rho}^0_i$, it is solved by the classical Picard iterations with underrelaxation involving intermediate step of \eqref{eq:Poisson_0}, \eqref{eq:bc_Poisson_0} solution for current densities guess.

For the sake of computational efficiency, the potentials $\overline{\varphi}^0_i$ are not evaluated ``on the fly" during the solution of the dynamic equations \eqref{eq:transport_average}. Instead, the potentials are first calculated on sufficiently fine grid in average densities space. Next, smooth interpolation is built based on the calculated values. Then, the interpolants are used while solving the dynamic problem.
\section{Density Functional Theory}
\label{sec:DFT}
Here, we describe in the detail thermodynamic model of electrolyte inside nanopores, which is based on Classical Density Functional Theory (c-DFT). The version of this approach developed for neutral molecules is able to take into account the influence of nanoscale geometrical constraints \cite{roth2010fundamental}. The confined fluid model can be extended to the electrolyte fluid  accounting for electrostatic correlations and external Coulomb field \cite{wang2011weighted}. We consider an open slit pore stored by neutral electrolyte mixture $\sum_{k=1}^nZ_k=0$ with known composition and chemical potentials $\{\mu\}_{k=1}^n$. Such confined system is described in terms of the Grand Canonical potential $\Omega$ and external field:

\begin{equation}
\label{eq:omega_potetial}
\Omega[\{\rho_i(\mathbf{r})\}]=F[\{\rho_i(\mathbf{r})\}]+\sum_{i=1}^n\int d\mathbf{r}\rho_i(\mathbf{r})\left(U_\text{ext,i}(\mathbf{r}) -\mu_i\right)
\end{equation}
where $U_\text{ext}$ is the external field acting on a fluid molecule, $\mu$ is the chemical potential. In the case of charged molecules, the external fields contains not only wall potential, but also Coulomb field contribution: 
\begin{equation}
\label{eq:ext_potential}
U_\text{ext,i}=U_\text{w,i}+U_\text{C,i}
\end{equation}
In our study, we use hard sphere potential to described non-electrostatic fluid-solid interactions:
\begin{equation}
\label{eq:hard_wall}
U_\text{w,i}(r)=\systeme{
\infty \quad \text{if} \quad r<d_i/2, 
0 \quad \text{if} \quad r>d_i/2}
\end{equation}
The slit pore geometry allows us to reduce the spatial density distribution $\rho_i(\mathbf{r})$ to 1D function $\rho_i(z)$ of the normal distance to the solid surface. The Helmholtz energy of ionic liquids can be written as the following:  
\begin{equation}
\label{eq:free_energy}
F=F_\text{id}+F_\text{hs}+F_\text{C}+F_\text{el}
\end{equation}
where $F_\text{id}$ is the ideal gas contribution; $F_\text{hs}$ is the hard sphere term accounting for ions excluded volume effects; $F_\text{C}$ is the Coulomb interaction; $F_\text{el}$ is the electric residual contribution. Here, only the ideal part is known exactly:
\begin{equation}
F_\text{id}=A k_\text{B}T\sum_{i=1}^n\int dz \rho_i(z)\log\left(\left[\Lambda^3\rho(z)\right]-1\right)
\end{equation}
the remaining terms define the excess part of the total Helmholtz free energy:
\begin{equation}
\label{eq:excess_free_energy}
F_\text{exc}=F_\text{hs}+F_\text{C}+F_\text{el}
\end{equation}

In accordance with the DFT approach, the equilibrium density distributions are defined by the following system:
\begin{equation}
\label{eq:DFT_condition}
\frac{\delta \Omega}{\delta \rho_i}=0, \quad i=1,...,n
\end{equation}
After substitution of expressions for the Helmholtz free energy \eqref{eq:free_energy} and the external potential \eqref{eq:ext_potential}, the conditions \eqref{eq:DFT_condition} has more explicit form:
\begin{equation}
\label{eq:master_eq_0}
\rho_i=\rho^{0}_i\exp\left[-\beta U_\text{C,i}- \beta U_\text{w,i}-\lambda_i\right]
\end{equation}
where $\rho^{0}_i$ is the bulk component density, $\lambda_i$ is the density derivative of the deviation of the excess terms from the bulk ones:
\begin{equation}
\lambda_i=\beta\frac{\delta(F_\text{exc}-F_\text{exc}^{0})}{\delta \rho_i}
\end{equation}
where $F_\text{exc}^{0}$ is the bulk excess free energy corresponding to the homogeneous mixture $\{\rho^0_k\}_{k=1}^n$.

The functional derivative of the Coulomb contribution has the following form:

\begin{equation}
\label{eq:Coulomb}
\beta \frac{\delta F_\text{C}}{\delta \rho_i}= Z_i l_\text{B} \sum_{j=1}^n Z_j\int d \mathbf{s} \frac{\rho_j(\mathbf{s})}{|\mathbf{s}-\mathbf{r}|}
\end{equation}
where $l_\text{B}=\beta e^2/4\pi \epsilon \epsilon_0$ is the Bjerrum length. The right hand of expression \eqref{eq:Coulomb} can be rewritten in terms of external Coulomb field $U_C$ and the mean electrostatic potential $\psi$:
\begin{equation}
\label{eq:el_relation}
Z_i e \psi-U_{C,i}=Z_i l_\text{B} \sum_{j=1}^n Z_j\int d \mathbf{s} \frac{\rho_j(\mathbf{s})}{|\mathbf{s}-\mathbf{r}|}
\end{equation}
Using expressions \eqref{eq:Coulomb}, \eqref{eq:el_relation} equation \eqref{eq:master_eq_0} can be rewritten as follows:
\begin{equation}
\label{eq:master_eq_1}
\rho_i=\rho^{(0)}_i\exp\left[- \beta U_\text{w,i} -\beta Z_i e \psi-\frac{\beta\delta}{A\delta \rho}\left(\Delta F_\text{hs}+\Delta F_\text{el}\right)\right]
\end{equation}
where symbol $\Delta$ means the difference between confined and bulk energies.  

The hard sphere contribution can be calculated using Fundamental Measure Theory \cite{roth2010fundamental} as:
\begin{equation}
\label{eq:HS}
\beta F_\text{hs}[\rho_1(z),...,\rho_n(z)]=A\int_0^H dz'\Phi(n_0,n_1,n_2,n_3,\mathbf{n}_{v1},\mathbf{n}_{v2})
\end{equation}
where the function $\Phi$ depends on the weighted densities $n_\alpha$ defined as:

\begin{align}
\label{eq:weighted}
n_0=\sum_{k=1}^n\frac{1}{d_i}\int_{z-R}^{z+R} \rho_k(z') dz' \nonumber \\
n_1=\frac{1}{2}\sum_{k=1}^n\int_{z-R}^{z+R} \rho_k(z') dz'\nonumber \\
n_2=\sum_{k=1}^n\pi d_i\int_{z-R}^{z+R} \rho_k(z') dz' \\
n_3=\frac{1}{2}\sum_{k=1}^n\int_{z-R}^{z+R} \left[d_k^2-(z-z')^2\right]\rho_k(z') dz' \nonumber \\
\mathbf{n}_{v1}=-\sum_{k=1}^n \frac{1}{d_k}\frac{\mathbf{z}}{z}\int_{z-R}^{z+R}dz'(z'-z)\rho(z') \nonumber \\
\mathbf{n}_{v2}=-2\pi\sum_{k=1}^n \frac{\mathbf{z}}{z}\int_{z-R}^{z+R}dz'(z'-z)\rho(z') \nonumber 
\end{align}We use one of the most popular version of the $\Phi$ defined as:
\begin{equation}
\label{eq:hs_density_energy}
\Phi=-n_0 \log(1-n_3)+(n_1 n_2 -\mathbf{n}_{v1}\mathbf{n}_{v2})/(1-n_3)+
1/(36\pi) (n_3 \log(1-n_3)+n_3^2/(1-n_3)^2)(n_2^3-3n_2 \mathbf{n}_{v2}^2)/n_3^3]
\end{equation}

Therefore, the functional derivative of the hard sphere contribution can be calculated by the following way:

\begin{equation}
\label{eq:hs_functional_derivative}
\frac{\beta\delta F_\text{hs}}{\delta_i \rho(z)}=A\int_0^H dz'\sum_\alpha\frac{\partial \Phi(n_\alpha)}{\partial n_\alpha}\frac{\delta n_\alpha}{\delta\rho_i}
\end{equation}

To calculate electrostatic term in equation \eqref{eq:master_eq_1}, we use the approach described in work \cite{wang2011weighted}:
\begin{equation}
\label{eq:EL}
\frac{\delta \Delta F_\text{el}}{\delta \rho_i}=-\sum_{k=1}^n\int d\mathbf{s}\bar{c}(\mathbf{r},\mathbf{s})\Delta\rho_k(\mathbf{s})
\end{equation}
where the weighted correlation function defined as
\begin{equation}
\label{eq:weighted_corr}
\bar{c}_{ki}(\mathbf{r},\mathbf{s})=\frac{\int c_{ki}(\mathbf{r}',\mathbf{s})f_{ki}(\mathbf{r}')}{\int d\mathbf{r}' f_{ki}(\mathbf{r}')}
\end{equation}
In our work, we use the FMT/WCA-k$^2$ approach that corresponds to the following expression of f-function:

\begin{equation}
\label{eq:f-function}
f_{ki}(\mathbf{r}')=\kappa^2(\mathbf{r}')\Theta(|\mathbf{r}-\mathbf{r}'|-d_{ki})
\end{equation}
where $d_{ki}=(d_k+d_i)/2 $  is the average diameter, $\kappa$ is the Debye parameter given by

\begin{equation}
\label{eq:Debye}
\kappa^2(\mathbf{r'})=4\pi l_\text{B}\sum_{k}^n\rho_k Z_k
\end{equation}
In accordance with work \cite{wang2011weighted}, here we use the approximated analytical expression for the weighted correlation function in terms of MSA solution:
\begin{equation}
\bar{c}_{ki}(\mathbf{r}',\mathbf{s})\simeq c^{MSA}_{ki}(|\mathbf{r}'-\mathbf{s}|)=\beta U_{ki}(r)\left[1-B_{ki}(\mathbf{r}')\frac{r}{d_{ki}}\right]^2\Theta(d_{ki}-r)
\end{equation}
where $r=|\mathbf{r}'-\mathbf{s}|$ is the scalar distance, $B_{ki}$ depends analytically  on Debye parameter $\kappa$ \eqref{eq:Debye} as
\begin{equation}
B_{ki}=\frac{1+\kappa d_{ki}-\sqrt{1+2\kappa d_{ki}}}{\kappa d_{ki}}
\end{equation}
Therefore, the weighted correlation function \eqref{eq:weighted_corr} in slit geometry $\rho(z)$ can be written as:

\begin{equation}
\label{eq:weighted_corr_1}
\bar{c}_{ki}(r)=\beta U_{ki}(r)\left[1-2B_{1,ki}(z) \frac{r}{d_{ki}}+B_{2,ki}\left(\frac{r}{d_{ki}}\right)\right]\Theta(d_{ij}-r)
\end{equation}
where
\begin{equation}
B_{m,ki}(z)=\frac{\int_{z-d_{ki}}^{z+d_{ki}} dz' B^m(z')\kappa^2(z')(d^2-(z-z')^2) }{\int_{z-d_{ki}}^{z+d_{ki}} dz' \kappa^2(z')(d^2-(z-z')^2)}
\end{equation}

\begin{figure*}[t!]
    \subfigure{
        \includegraphics[width=0.31\textwidth]{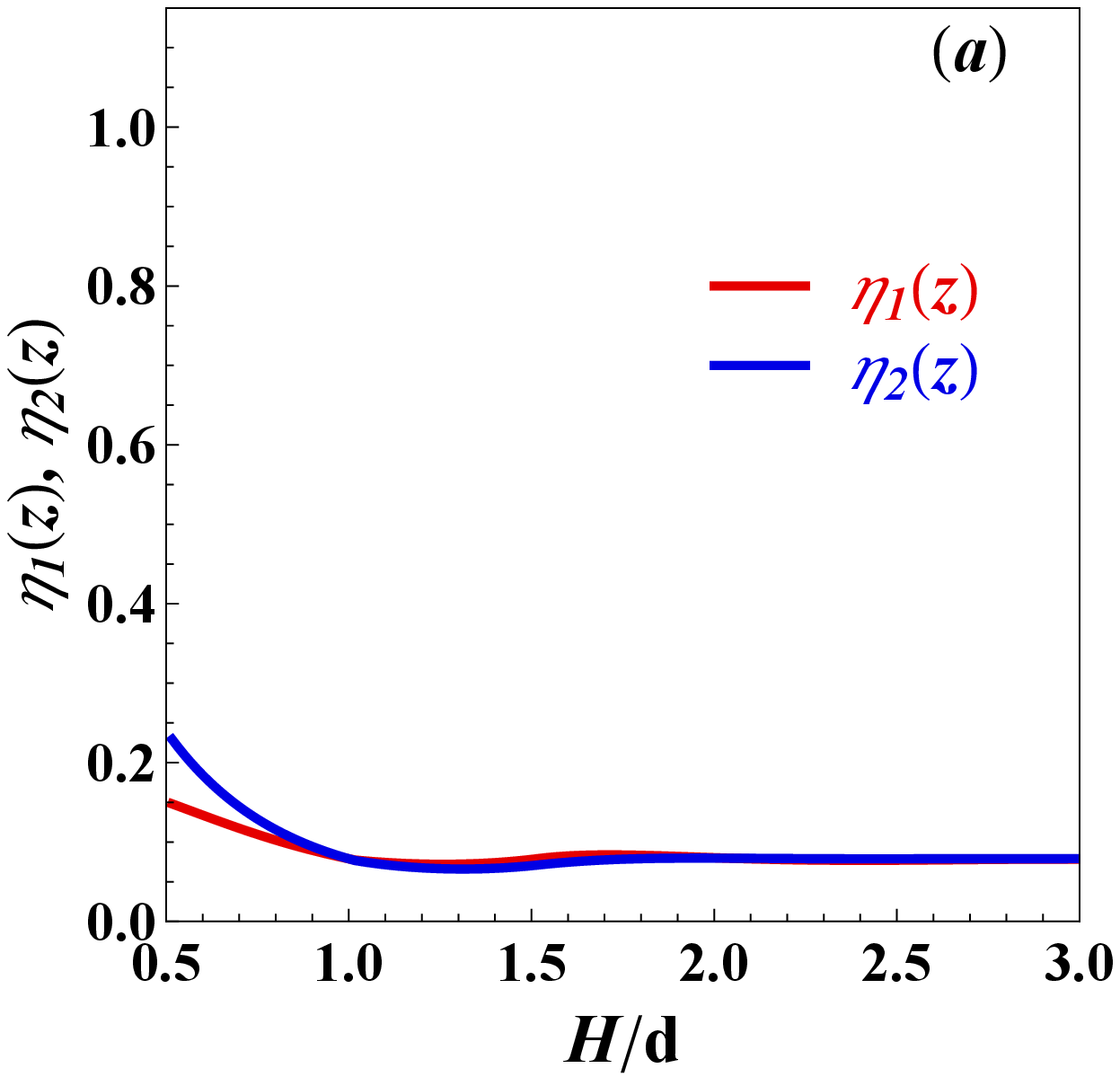}
        \label{fig:density_U05}
    }
    \subfigure{
        \includegraphics[width=0.31\textwidth]{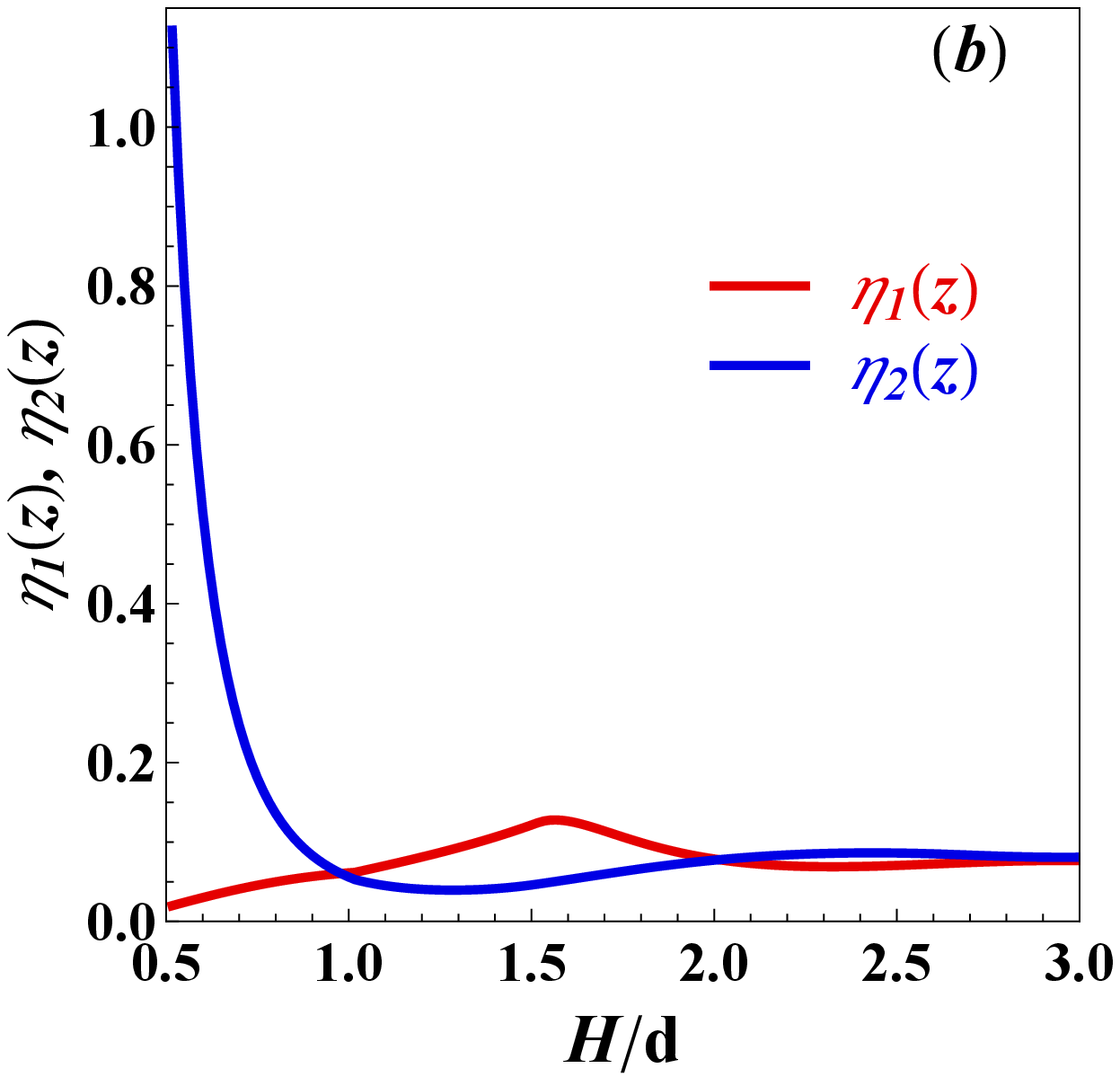}
        \label{fig:density_U5}
    }
    \subfigure{
        \includegraphics[width=0.31\textwidth]{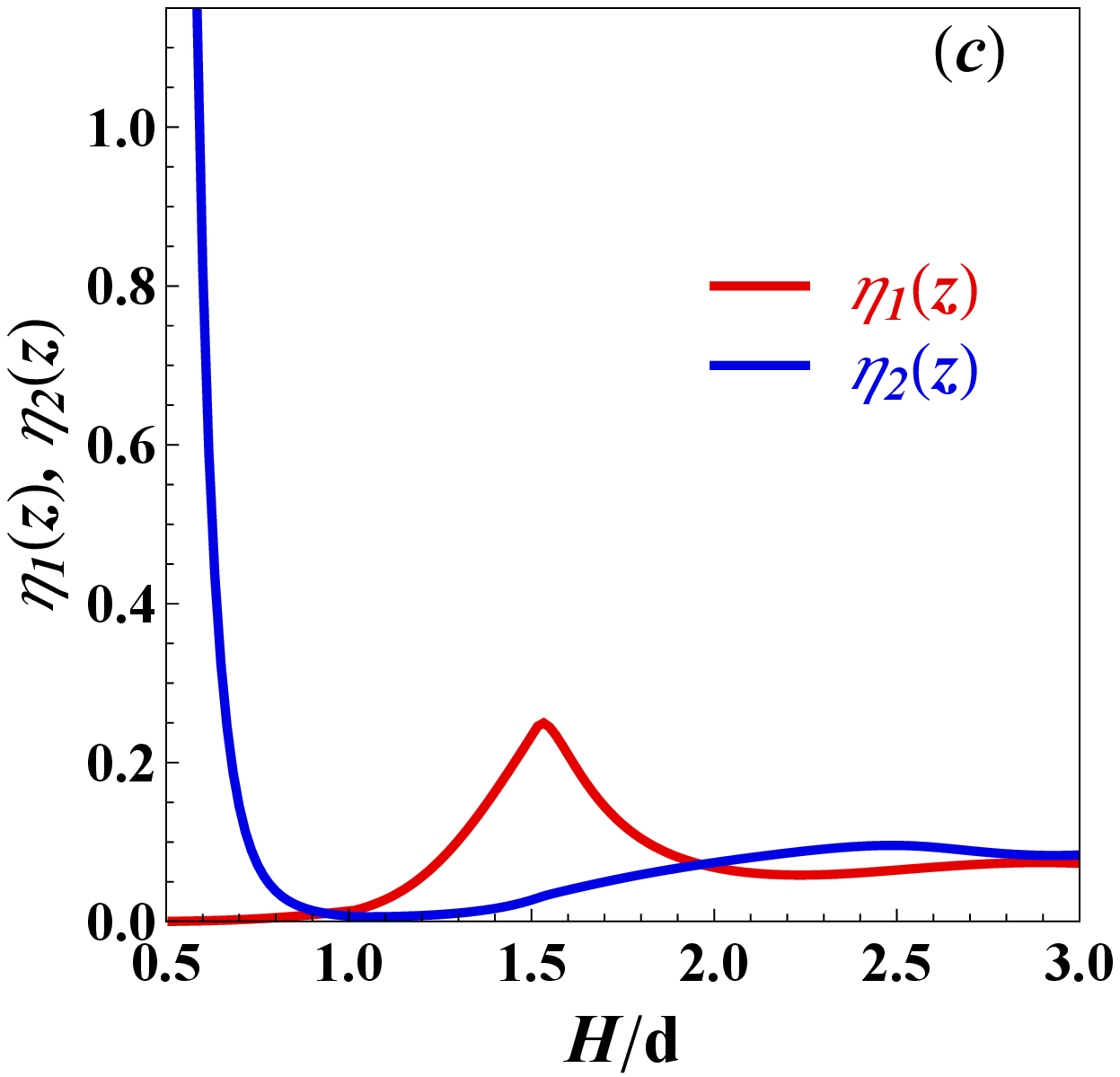}
        \label{fig:density_U20}
    }
    \caption{The dimensionless density distributions of positive $\eta_1(z)=v\rho_1(z)$ and negative $\eta_2(z)=v\rho_2(z)$ electrolyte's components inside pore $H^*=6$ at various applied potentials $U^*=0.5$ (a), $U^*=5$ (b) and $U^*=20$ (c). These calculations correspond to the bulk density $\rho^*_{0,1}=\rho^*_{0,2}=0.15$ at the temperature $T^*=0.15$.} 
    \label{fig:density_profiles}
\end{figure*}

In our work, we use the same system parameters as in work \cite{pizio2012electric} corresponding to the dimensional temperature $T^*=d/l_\text{B}=0.15$. The characteristic dependence of the density distribution profiles on applied potential is shown in Fig.~\ref{fig:density_profiles}. As shown in \cite{pizio2012electric}, such behaviour of co- and counter-ions results in the oscillating capacity properties.

\section{Connection with Transition Line Model (TLM)}
Here we supplement the analysis of the first exponential regime's time scale 
\begin{equation}
\label{eq:first_time_scale}
\tau_1^* =\frac{4}{\pi^2}\frac{Q^*_{\infty}}{\overline{\rho}^*_{\infty}U^*H^*}
\end{equation}
by another connection with the result of TLM. In \cite{bi2020molecular} TLM solution was used to describe the charging inside nanoporous Metal Organic Frameworks (MOFs) electrodes. The following equation for the net pore charge after jump-wise application of constant potential was obtained:
\begin{equation}
    \label{eq:TLM_charge}
    \frac{Q(t)}{Q_\infty}=1-\frac{2}{\pi^2}\sum_{n=0}^\infty\frac{\exp\left[-\pi^2(n+1/2)^2(l/L)^2 t/\tau\right]}{(n+1/2)^2}.
\end{equation}
Here $\tau=C_a l/\sigma$ is the intrinsic relaxation time, $C_a$ is the areal capacity of the pore, $l$ is the pore volume divided by its surface area, $\sigma$ is the ionic conductivity inside the pore. Please note, that the exponent in \eqref{eq:TLM_charge} differs from the given in the original paper by the factor of 4, because the symmetric pore connected to bulk electrolyte at both ends and effectively composed of two closed-end pores considered here was studied in \cite{bi2020molecular}. 

The series in the right hand of expression \eqref{eq:TLM_charge} can be approximated by the leading term $n=0$, that result in the following time-scale of TLM model:
\begin{equation}
    \label{eq:TLM_scale}
    \tau_\text{aTLM}=\frac{4}{\pi^2}\frac{C_a L^2}{\sigma l}.
\end{equation}

For the slit pore geometry $l \approx H / 2$. Using the notation adopted in this paper the areal capacitance can be expressed as $C_a  = Q_\infty / 2U$. Finally, the ionic conductivity inside the pore can be expressed in terms of the electrical mobility of ions $m = e D / k_B T$ and ions density $\overline{\rho}_\infty$ as $\sigma = e \overline{\rho}_\infty m$. Substitution of these expressions to \eqref{eq:TLM_scale} gives
\begin{equation*}
\tau_\text{aTLM}=\frac{4}{\pi^2} \frac{Q_\infty k_B T}{e^2 \overline{\rho}_\infty U H} \frac{L^2}{D}.
\end{equation*}
Introducing dimensionless variables $H^*=H/d$, $U^*=e U/k_B T$, $Q^*=Q d^2/e$, $\rho^* = \rho d^3$ and $t^* = t D / L^2$ used in the main document into the latter expression one can readily obtain that $\tau^*_\text{aTLM} = \tau^*_1$, i.e. TLM time scale is equivalent to the time given by \eqref{eq:first_time_scale}.

\end{document}